# The Effectiveness of Social Media Engagement Strategy on Disaster Fundraising

Vivek Velivela
School of Computer Science, University of Technology Sydney, Sydney, Australia
Velivelavivek2000@gmail.com

Chahat Raj
School of Computer Science, University of Technology Sydney, Sydney, Australia
Chahatraj58@gmail.com

Muhammad Salman Tiwana
School of Computer Science, University of Technology Sydney, Sydney, Australia
salmantiwana@ymail.com

Raj Prasanna
School of Psychology, Messey University, Palmerston North, New Zealand
r.prasanna@massey.ac.nz

Mahendra Samarawickrama
Australian Red Cross, Sydney, Australia
samarawickrama@gmail.com

Mukesh Prasad
School of Computer Science, University of Technology Sydney, Sydney, Australia
Mukesh.prasad@uts.edu.au

**ABSTRACT**

Social media has been a powerful tool and integral part of communication, especially during natural disasters. Social media platforms help nonprofits in effective disaster management by disseminating crucial information to various communities at the earliest. Besides spreading information to every corner of the world, various platforms incorporate many features that give access to host online fundraising events, process online donations, etc. The current literature lacks the theoretical structure investigating the correlation between social media engagement and crisis management. Large nonprofit organisations like the Australian Red Cross have upscaled their operations to help nearly 6,000 bushfire survivors through various grants and helped 21,563 people with psychological support and other assistance through their recovery program (Australian Red Cross, 2021). This paper considers the case of bushfires in Australia 2019-2020 to inspect the role of social media in escalating fundraising via analysing the donation data of the Australian Red Cross from October 2019 - March 2020 and analysing the level of public interaction with their Facebook page and its content in the same period.

**INTRODUCTION**

The Bushfires in Australia (2019-2020) are considered enormous scale destruction in the modern record in New South Wales.  Its significance is due to the record number of deaths and longevity from Nov-2019 to Mar-2020 (Smith, 2021). The bushfires have destroyed 3,094 houses and burnt over 17 million hectares of land across various states in Australia (Smith, 2021). The widespread fires have perished over 1 billion animals causing wildlife devastation (Green, 2020). Nonprofit organisations like Australian Red Cross have played a crucial role with certain governments in minimising the damage caused by this natural disaster by mobilising a certain amount of financial and human resources to help people or communities in need with a presence in metropolitan, rural and remote areas (Tanveer et al., 2020). The Australian Red Cross is a part of the world's most significant humanitarian movement, operating in over 190 countries and has over 17 million volunteers worldwide, and has raised the highest donations with more than a quarter-billion in the Australian bushfire crisis season. Recently, social media had an important role in promoting and amplifying the community's passion for climate change (Schweinsberg et al., 2020). Similarly, nonprofit organisations have been using social media to spread important events like earth day to raise awareness of critical concerns like global warming. As social media evolves, nonprofit organisations have more flexibility to conduct online events that can positively impact society (Tsai et al., 2019). Large nonprofit organisations like the Australian Red Cross have a significant social media presence.

During the 2019-2020 bushfires, huge social media interaction was observed where internet users communicated and expressed their emotions about the disaster. Social media is emerging as an effective tool to generate social awareness, spread official advisories, and raise relief funds (Adhikari et al., 2022). The Australian Red Cross employed its Facebook page to generate awareness among social media users and raise donations by indulging users in posts and videos (Rajora et al., 2018). The donation amount rose during the bushfire period, which seemingly influenced social media interaction. However, no study has analysed such a relationship to establish its concreteness.

This paper provides descriptive-analytical insights of the Australian Red Cross Facebook post interactions and donation patterns in bushfire season and discusses the role of social media during Australian bushfires (2019-2020).

**RELATED WORK**

Social media engagement has established a crucial role in emergencies and disaster response. However, research focusing on such technological employment is relatively limited (Haworth & Bruce, 2015). The widespread increase in different diseases after the disaster hits also causes alarming situation (Kabade et al., 2021). Government and non-government organisations communicate through social media to target large populations, strengthening disaster response capabilities (Poblet et al., 2018). The disaster management cycle involves four critical steps: prevention, preparedness, response, and recovery, which are eminently supported by social media platforms. Houston et al. developed an architecture depicting various dimensions of social media usage towards disaster management (Houston et al., 2015). The framework allows the stakeholders to undergo the disaster management cycle by encompassing preparedness, warnings and advisories, and disaster predictions. Organisations responsible for incident response actively engage social media users to communicate disaster warnings, directions, and advice (Erfani et al., 2016; Whitelaw & Henson, 2014). Haworth et al. established the importance of social platforms and Volunteered Geographic Information (VGI) sharing through a survey of 154 Tasmanian participants demonstrating significant potential of bushfire communication through online platforms (Haworth et al., 2015). A system proposed for classification of data extracted from disaster crises management websites by implementing some classification models. Implementation was conducted in two modules one for extraction of data and other for classification of data related to disaster or not. Domala et al. used bags of words and TF-IDF vectors as a feature extraction technique along with the well-known classifiers named as linear and logistic regression to achieve best accuracy (Domala et al., 2020). Perera et al. proposed a real-time approach using satellite imagery and remote sensing to proliferate bushfire-related warnings to the general public using social media such as Facebook and Twitter (Perera et al., 2021). The proposed framework allows social media users to contribute geographical information through images generating advanced bushfire warnings. The indulgence of social media users aims to disseminate the maximum possible information, warnings, and preventive methods and reach rural and semi-rural communities to mitigate bushfire impacts. (Abedin & Babar, 2018) emphasised the usefulness of microblogging on the Twitter platform, suggesting that the tweets were more informative than directive. They also characterise Twitter usage by formal institutions and non-institutional volunteers during a natural calamity concluding that individual users greatly influenced information dissemination during Australian bushfires. Anikeeva et al. suggested that Twitter can be used as a broadcasting tool to deliver frequent updates to the citizens, whereas Facebook can be used for establishing online communities, for instance, the Australian Red Cross page (Anikeeva et al., 2015). During the disaster (Sufi et al., 2022) extracted location-oriented data and implemented Artificial Intelligence (AI) and Natural Language Processing (NLP) based algorithm for sentiment analysis. They designed an approach having vast knowledge of responses collected from social media related to disaster in 39 different languages by using AL and NLP for analysis and detection of anomaly, regression and applied Getis Ord Gi∗ algorithms. It was found that 70% of tweets collected were live location-oriented from different location effected by disaster. The accuracy calculated at last was about 97% on the data set. Roshan et al. investigated the use of Facebook and Twitter for crisis communication by similar organisations by performing qualitative content analysis and gain essential insights into usage patterns and motives (Roshan et al., 2016). They concluded that organisations are not fully utilising the social media potential for communication during natural disasters.

**MATERIALS AND METHODS**

**Data**

We have extracted the time series bushfire donation data of the Australian Red Cross from 2011-2021 and social media engagement time series data of the Australian Red Cross Facebook page from October 2019-March 2020. Out of the large donation dataset, we only used data from October 2019- March 2020 aggregated by each month to analyse the dependence of social media factors on the donation amounts. After receiving the data, we have removed personally identifiable information (PII), geographical and additional data from the datasets. Data with no relevance is identified and not used for analysis purposes. The features analysed in this study from the aggregation of donation and social media data are listed in Table 1.

Table 1: List of features in the dataset and their explanation

| Abv. | Feature | Explanation |
|---|---|---|
| DA | Donation Amount | The total amount of donation collected through fundraising |
| F1 | Likes | Total number of likes on the posts of Australian Red Cross Facebook page |
| F2 | Shares | Total number of shares of all posts on the page during the study period |
| F3 | Comments | Total number of comments on all the posts generated during the study period on the Facebook page |
| F4 | Reacts | Total number of reactions (like, love, care, haha, wow, sad, angry) obtained on all the posts |
| F5 | Impressions | Total number of times the posts from the Australian Red Cross page was displayed on users' timeline (Reach) |
| F6 | Seconds viewed | The number of seconds a video was viewed |
| F7 | Duration (Sec) | The total duration of a video in seconds |
| F8 | 3-second video views | The total number of views where the users watched a video for 3 seconds |
| F9 | 60-second video views | The total number of views where the users watched a video for 60 seconds |
| F10 | Averaged seconds viewed | The average number of seconds for which the users watched a video |
| F11 | Unique 60-second video views | The total number of views from unique users who watched the video for 60 seconds |
| F12 | 60-second views from Recommendations | The total number of views on the videos where a user watched the video upon recommendation for 60 seconds |
| F13 | 60-second views from Shares | The total number of views on the videos where a user watched the video shared by a friend for 60 seconds |
| F14 | 60-second views from Followers | The total number of views on the videos where a follower of the Australian Red Cross page watched the video for 60 seconds |
| F15 | Seconds viewed from Recommendations | The number of seconds a video was viewed upon recommendation |
| F16 | Seconds viewed from Shares | The number of seconds a video was viewed when shared by a Facebook friend |
| F17 | Seconds viewed from Followers | The number of seconds the page's followers viewed a video |
| F18 | Average Seconds viewed from Recommendations | The average duration a video was viewed upon recommendation (in seconds) |
| F19 | Average Seconds viewed from Shares | The average duration a video was viewed when shared by a Facebook friend (in seconds) |
| F20 | Average Seconds viewed from Followers | The average duration a video was viewed by the page's followers (in seconds) |

**Methodology**

Social media is conducive to managing crises in terms of issue monitoring, framing crises, and reinforcing relationships with stakeholders while strengthening relationships with the online community in challenging times (Jiang et al., 2016). According to this study, liking a post shows the user interest in a particular post. It builds a connection with other users who also like the same post on social media, which gives a sense of community that shares similar interests. Facebook updates the user network when a user likes a page or donates to a fundraiser; that particular action acts as a user endorsement and acceptance of that specific brand or organisation (Lee, 2021). This user endorsement might also lead to the user network engagement or views of that organization (Sawhney et al., 2019).

User engagement is the critical metric for a digital organisation (Aldous et al., 2019). According to this study, there is four user engagement with level 1 as views, level 2 as likes, level 3 as shares and comments, and level 4 as external postings. Each level indicates some user actions representing a certain measure of engagement with a lower to higher level as most basic to advanced engagement (Agarwal et al., 2020). Views indicate a more private level of social interaction where the longer a user sees a post, the more interested the user is in that post. Liking, commenting, and sharing is a public expression that can be visible to a user's network (Aldous et al., 2019). views, likes, shares, and comments can be considered primary and standard metrics to determine user interaction with a page or a post (Raj et al., 2021). Reach is an attribute used in this analysis that defines the number of people who read any content from a page or about a page (Facebook, 2021). The impression is another attribute used in this analysis that defines the number of times the content from a page has entered a person's screen (Facebook, 2021). We have considered five attributes for social media engagement: likes, comments, shares, reach, and impressions. Reach, and impressions are very close to being considered views, but these attributes are different in their ways. These five attributes have quantitative potential to show the degree of user engagement in the Australian bushfire crisis, which can help us correlate with the donation data to understand some patterns between donation and user engagement.

To elevate user engagement, Australian Red Cross Facebook page regularly publishes videos on their page. These videos aim to impact users, raise their concerns about disasters and generate fundraising significantly. We extract these video analytics to understand how users respond to such social media posts and how such interaction influences disaster fundraising.

The analysis procedure primarily involves identifying the correlation between the donation amount and social media engagement factors. We plot a confusion matrix for all the features in the aggregated dataset. The implementation is carried out on Google Colab using Python 3. Google colab also known as Collaboratory is a free online service, provides a platform for performing different operations by using machine learning and artificial intelligence techniques for research education. It is based on Jupiter notebook where users can do some programming and run the code. Collaboratory also provide python 2 and 3 runtime pre-defined libraries service such as TensorFlow, MxNet, and PyTorch to perform various tasks. After completing the task data file can be saved by exporting to google drive or system hard drive and Google cloud is hosting the Collaboratory service (Carneiro et al., 2018). The inferences yielded through the correlation matrix were considerably few, following which we plotted the graphs for all the social engagement features with respect to the donation amounts in a monthly fashion. The results obtained and their analyses are discussed in the next section.

**RESULTS AND ANALYSIS**

Australian Red Cross has an extensive volunteer and donor base across various parts of Australia (Smith, 2021). The extracted bushfire donation data from mid-October 2019 to March 2020 is shown in figure 1. We observe a gradual increase in donations from the mid of November 2019. Aggregated and average amount donated per month from October 2019 to March 2020 is shown in Figure 2 and Figure 3. The minimum amount donated in this period is AUD 0.01, whereas the maximum amount donated in this period is approximately 7 Million AUD. The total amount donated from 10-2019 to 03-2020 is approximately 200 Million AUD.

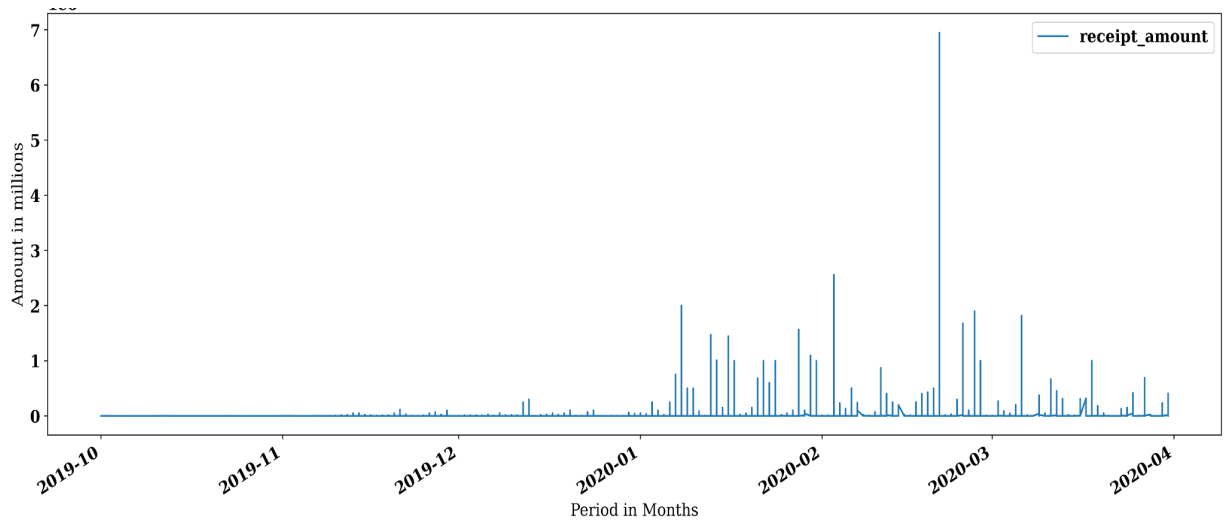

Figure 1: Donation amount ranging from October 2019 to March 2020 (Australian Bushfire crisis period)

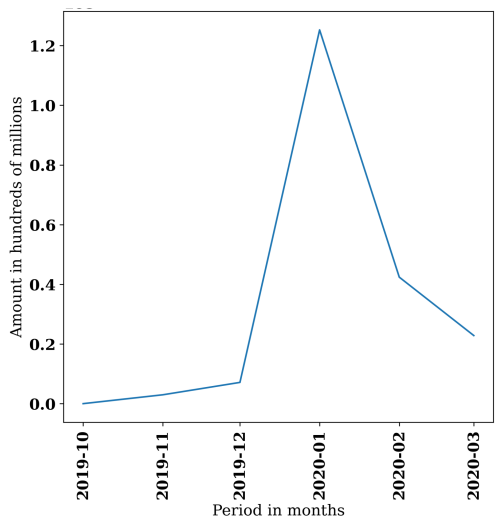

Figure 2: Monthly aggregated donation amount

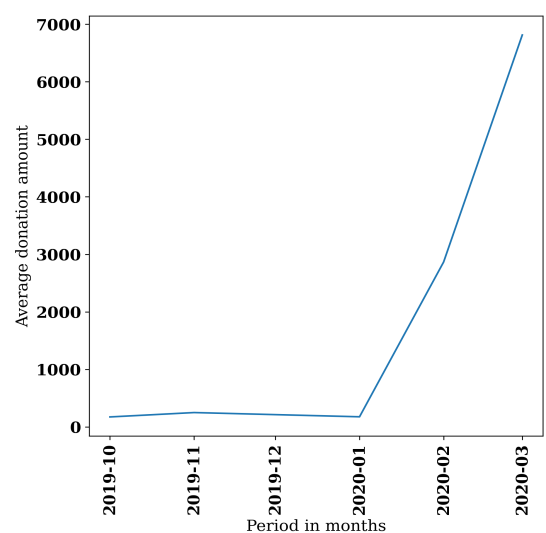

Figure 3: Monthly average donation amount

Australian Red Cross has a significant social media presence with about 253,000 followers, and 248,000 page likes on Facebook. Along with several followers, the Australian Red Cross Facebook consists of various types of content in the form of photos, videos, groups, fundraisers, and online events. We have extracted user engagement data from the Australian Red Cross Facebook Page from Oct-2019 to Mar-2020, including likes, comments, shares, Reach, Impressions and Seconds views of any content originating from the Australian Red Cross Facebook page or the page itself. Likes, comments, and shares of the Australian Red Cross page and its contents are shown in Figure 4. Reach, Impressions of the page and its content are shown in Figure 5.

From Figures 4 and 5, we can see that the user engagement has been steady from Oct 2019 to Nov 2019, whereas the spike of user engagement began in Nov 2019 lasted until Jan 2020. Although the intensity of the user engagement is different among likes, comments, shares, reach, impressions, and seconds viewed, they all show similar trends in that period. We also have extracted a correlation matrix among social media engagement attributes, as shown in Figure 6. We have used Pearson correlation to extract any attribute dependencies. We can see that there is a high correlation among various social media engagement attributes.

Figure 7 illustrates a correlation matrix for all the features employed in the exploratory analysis. It demonstrates a positive relationship between the donation amount with average seconds viewed, 60-second views from followers, seconds viewed from recommendations, followers, average seconds viewed from recommendations, shares, and followers. The highest correlation displayed is 96% between the donation amount and seconds viewed from the recommendation. This extreme value establishes a conclusive effect of social media recommendations. People are more likely to watch a video upon recommendations from their friends and acquaintances. Moreover, such recommendations are highly capable of convincing people to contribute towards disaster fundraising. 76% correlation observed between the donation amount and the average seconds viewed from recommendations reinforces this relationship. Another important conclusion drawn from the correlation matrix highlights that the video views received from the Australian Red Cross Facebook page followers drew massive donation amounts, with a positive correlation of 84%. Followers of a page or profile represent a concern to the cause and the community. Social media videos are more likely to influence the immediate followers of a page. Seconds viewed from followers also hold a 62% correlation with the donation amount. Similarly, a 63% correlation is observed between the donation amount and the average seconds viewed from shares, implying that the views generated upon sharing a Facebook post also significantly increased the donation amount. However, the correlation matrix failed to display any association of the donation amount with the rest of the features.

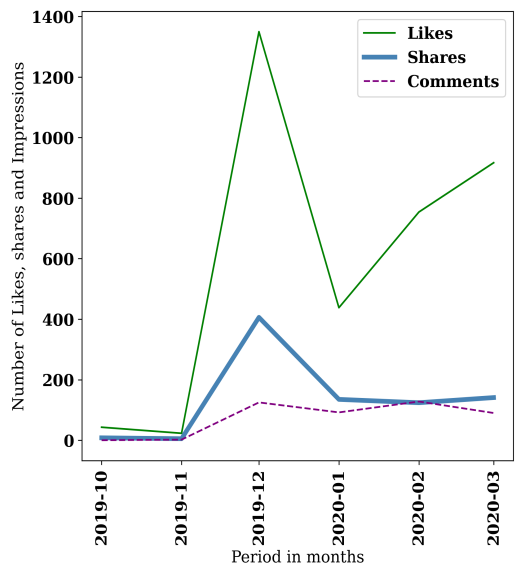

Figure 4: Likes, shares, comments in bushfire period

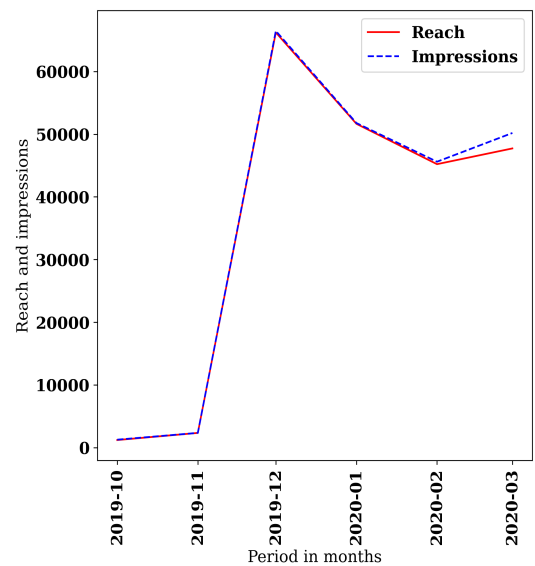

Figure 5: Reach and impressions in Bushfire period

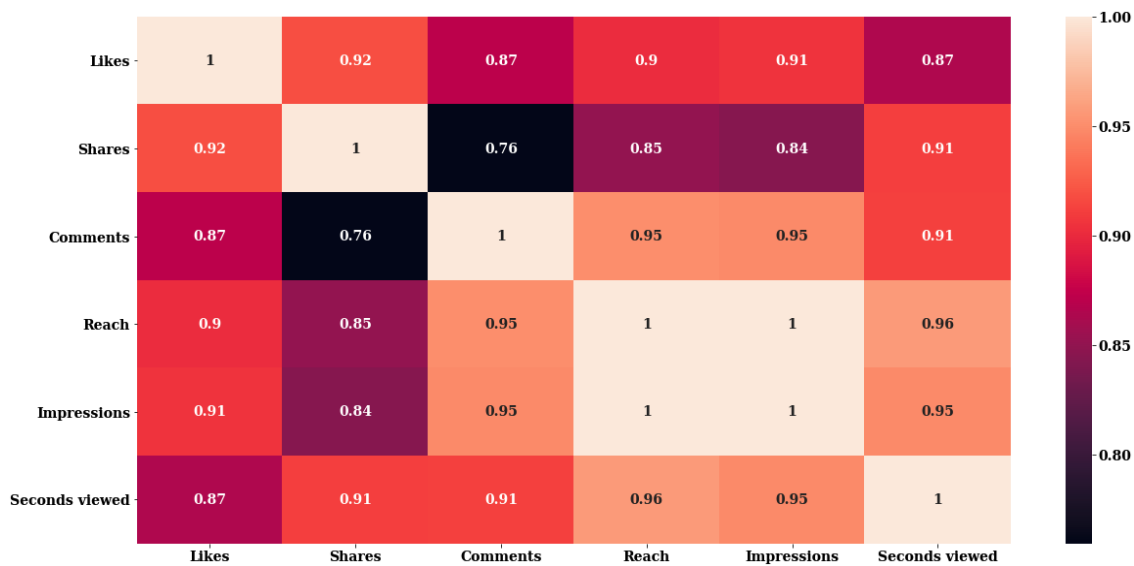

Figure 6: Correlation matrix among social media engagement attributes

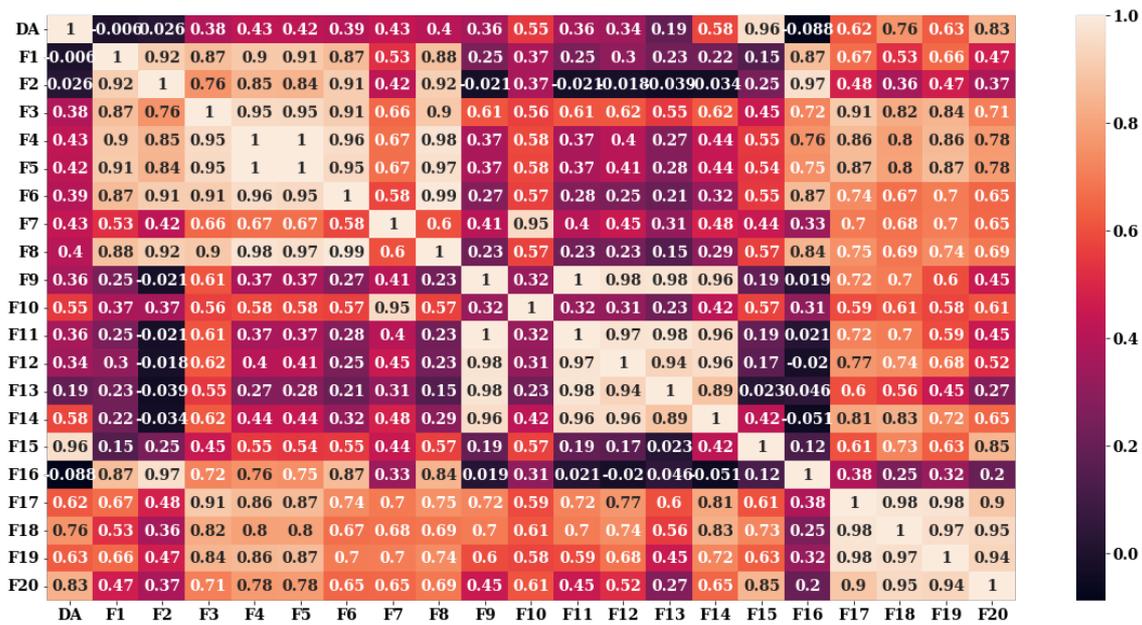

**Figure 7: Correlation matrix for social media engagement and video features with the donation amount**

Figure 8 demonstrates a normalised plot observing the donation revenue trends concerning user engagement on social media. The October to November phase during 2019 shows very gradual progress in the engagement features. The engagement escalates during November when all of the social media features, likes, shares, comments, reach, and impressions are observed to rise steeply till the peak. The beginning of December, however, decrements the engagement differently for all engagements. Likes and impressions demonstrate an approximate 65% decline, whereas shares, comments, and reach have reduced by ~20-25% since November. Analysing the donation amount alongside, it gradually increased between October and December, but a boom is demonstrated with a similar trend as the engagement factors during the December month. Following the graph, it is observable that the increment in engagement has a delayed, but direct impact on the donation amount as graphs demonstrate a similar slope for all the trendlines during this period. After reaching the peak donation amount in January 2020, it steadily reduces, following the trend of likes and impressions until February 2020. Post-February, the engagement factors demonstrate an irregular rise and fall, maintaining an engagement above ~35%. However, the donation amount steadily declines after this period.

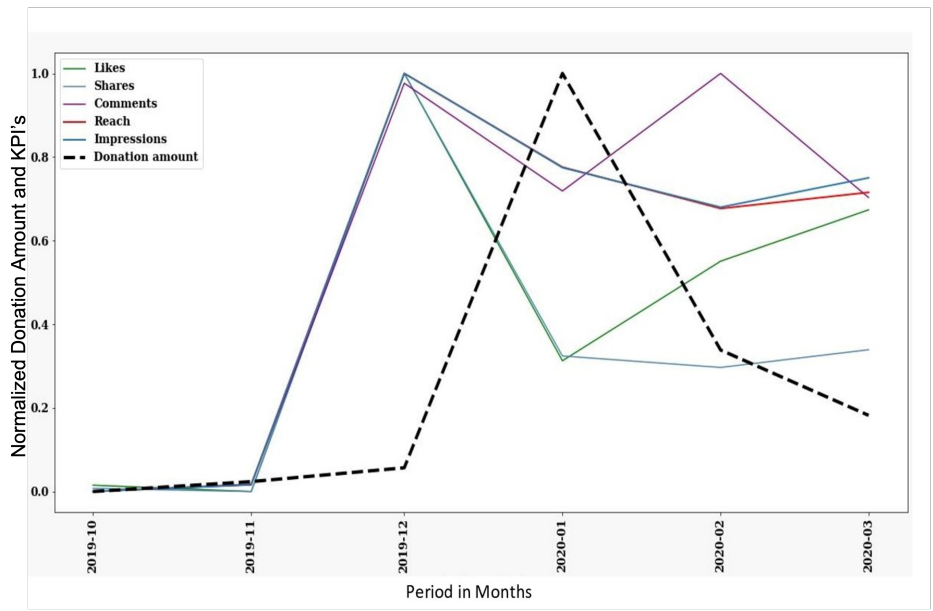

**Figure 8: Donation amount with respect to likes, shares, comments, reach, and impressions**

Next, we plot the donation amount with respect to the duration (Figure 9) and the average duration (Figure 10). The videos were watched due to unique factors- recommendations, shares, and followers. The video-viewing duration has risen throughout November 2019, with the steepest and maximum rise in the duration watched by shares. People who encounter a shared post in their news feed appear to watch the videos longer than follower or recommendation views. Followers watched the videos for an above-average duration (nearly above ~65% of the video's length). However, recommended videos were watched for a shorter duration of approximately 25% of the video's length. During December, the watching duration rose for views from recommendations and followers but experienced a fall for shared videos. On the whole, the donation trendline follows the pattern of watching duration from shares, experiencing similar but delayed peaks and lows.

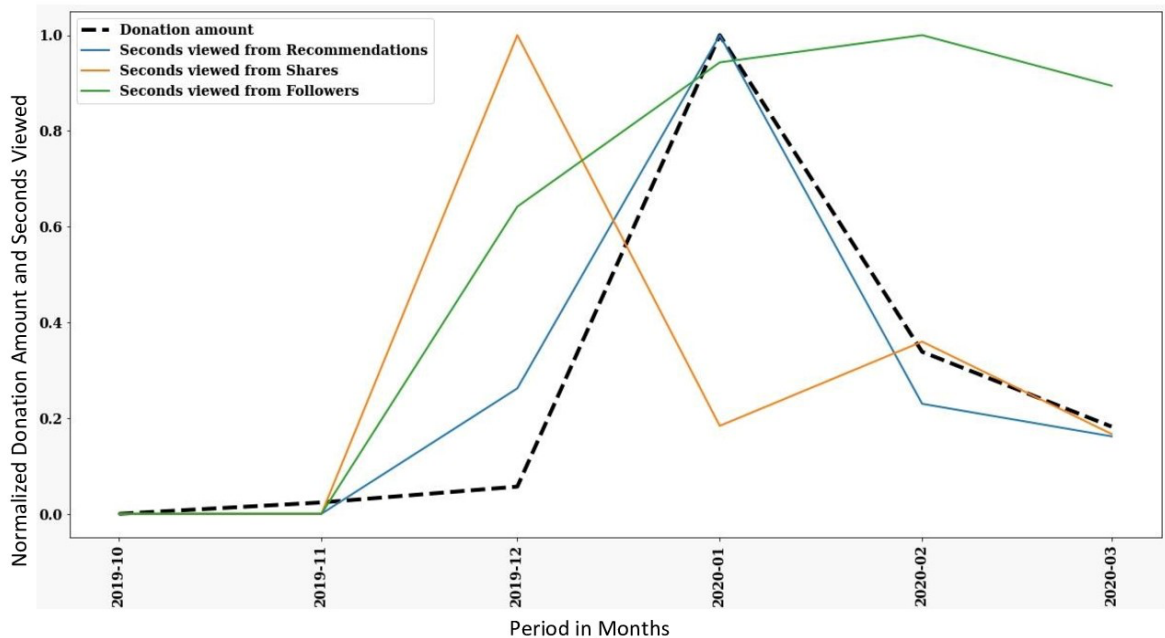

**Figure 9: Donation amount with respect to seconds viewed from recommendations, shares, and followers**

Averaging the watching duration from these views brought the trendlines closer for views from recommendations, followers, and shares, as illustrated in Figure 10. The average video-viewing duration continually rises for two months during November and December 2019, after which they experience a decline. The donation amount seemingly rises more steeply a month later. The spike is observed in average video viewing duration. This reestablishes the finding that video viewing had a direct but delayed response in disaster fundraising.

Figure 11 plots the donation amount against five key factors: video duration, video seconds viewed, 3-second views, 60-second views, and average seconds viewed. The donation amount rise during December 2019 is likely to be the effect of video duration. It is observed that increasing the length of the video has resulted in higher donation amounts. A similar trend is observed by the number of seconds viewed by social media users, the average number of seconds viewed, and the number of 3-second video views. All of these show a delayed impact, causing the donation amount to rise during December month. However, the rise in the number of viewers who watched the video for 60 seconds impacts the donation amount instantly. The number of views rises to nearly 30% while the donation amount peaks (~100%). Although during January, when the donation amounts begin to decline, the 60-second views are still rising, meaning that users are still engaging into watching the Australian Red Cross page videos, it is not generating any fundraising. Similar to other video viewing patterns and the donation amount, the 60-second views fall gradually after January.

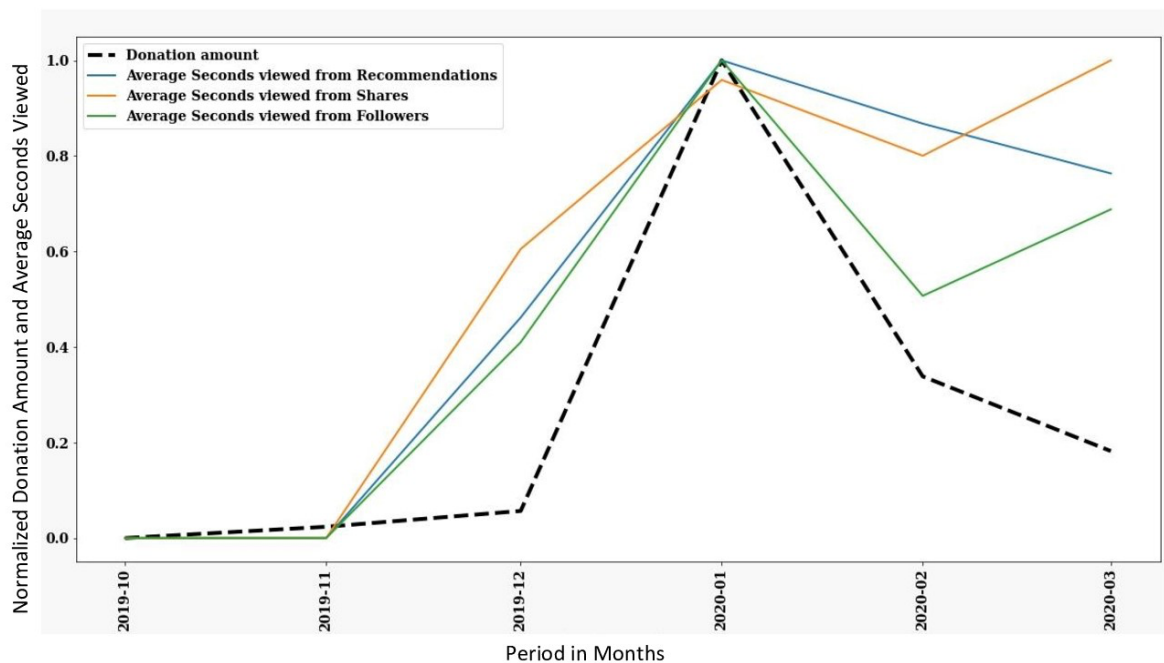

Figure 10: Donation amount w.r.t. average seconds video viewed from recommendations, shares, and followers

Figure 12 further breaks down the trend of 60-second video views into unique factors. Since 60-second views displayed an immediate response towards fundraising, it is essential to consider the constitution of these views. It is observed that the most views of the most prolonged duration were obtained from the page followers. Thus, highest engagement and impact were observed amongst the page followers, followed by the views from recommendations, unique views, and views from shares. The trendlines for unique video views and the views from recommendations move very close to each other, suggesting that the viewers who watched videos from recommendations were mostly unique viewers, with a marginal percentage of them being the existing followers of the page. The number of views from shares is the least among the 60-second views suggesting that sharing a video on a user's timeline brought substantially lower views and donations. During January 2020, the video views continued to rise, but the donation amount experienced a fall thereafter even if the number of views kept rising. Altogether, a drop is experienced during February 2020, suggesting that the users were uninterested or concerned after a particular time. This could be the case where the post experienced maximum visibility, and thereafter it was circulating among the same people who already viewed the videos or fundraised towards the cause.

**Key Findings**

1. Social media engagement displayed a direct but delayed relationship with the donation amount during the Australian bushfires.
2. The donation amount displays a positive correlation with the number of views on the Australian Red Cross Facebook page videos.
3. The six social media engagement attributes, namely, likes, shares, comments, reach, impressions, and seconds viewed, demonstrated a high positive correlation (>75%) amongst each other.
4. The rise in social media engagement through likes, shares, comments, reach, and impressions during November 2019 caused a similar increment in the donation amount during December 2019.
5. The decrement in social media engagement in December 2019 caused a similar decline in donation amount during January 2020.
6. As users were watching a video for longer durations, the donation amount also hiked.
7. Recommended videos gained longer views from fellow users and reached a peak by November end.
8. The donation amount increased with the increase in video duration.
9. Videos watched for 60 seconds influenced donation amounts in the same month, immediately impacting the viewers.
10. The earliest video views from followers were to ascend, followed by recommendations, unique views, and views from shares.

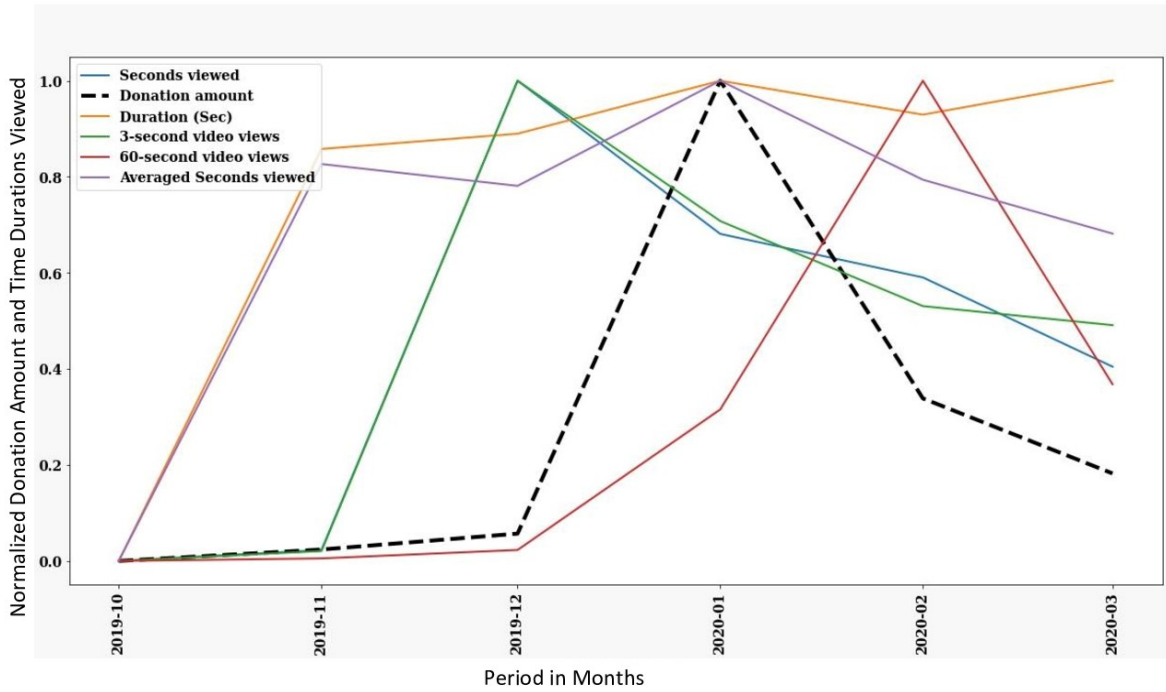

Figure 11: Donation amount with respect to video duration, video seconds viewed, 3-second views, 60-second views, and average seconds viewed

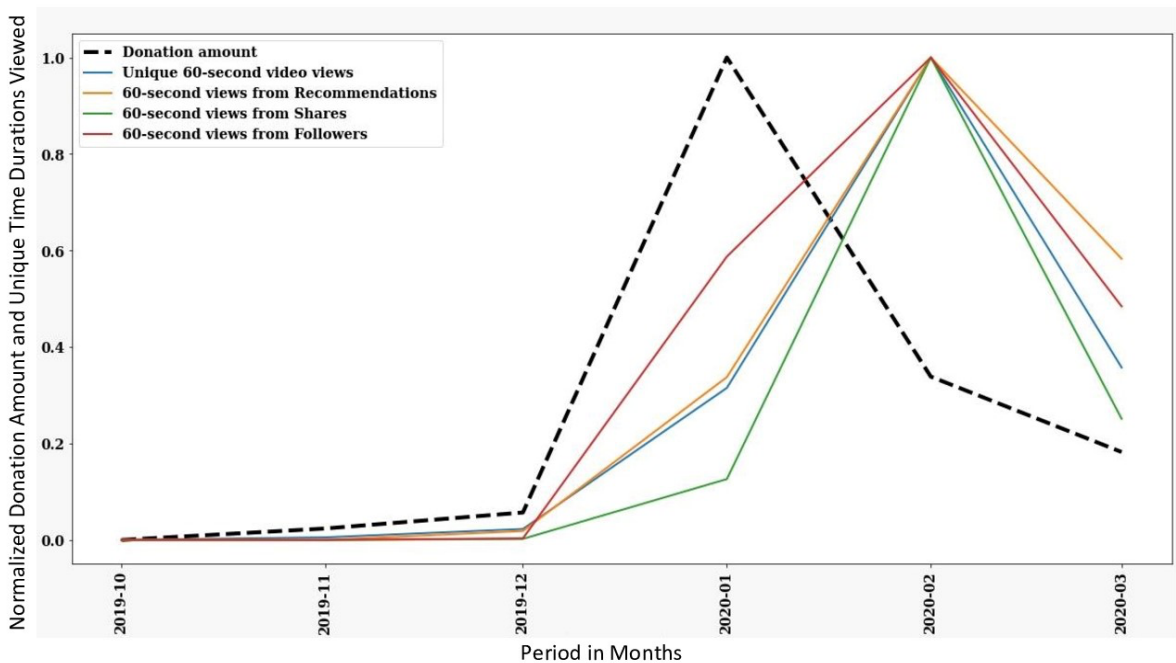

Figure 12: Donation amount with respect to 60-seconds unique video views, views from recommendations, shares, and followers

## CONCLUSION

In this exploratory research, we have investigated novel and unique Australian Red Cross bushfire donation trends in accordance with social media engagement. This analysis offers an opportunity to explore disaster donations and Facebook interactions in a relevant way. The social media engagement attributes (likes, shares, comments, reach, and impressions) are highly correlated. This suggests that user involvement is high with respect to all levels of engagement. The correlation between likes, shares, and comments suggests that more than private engagement, user advocacy via sharing and commenting posts in that period might have played an essential role in the revenue growth of donation. We observed a similar trend in donation revenue and social media engagement at different times of the year. The spike in user engagement before donations also suggests that social media's level of user engagement might have acted as a catalyst in generating the donation revenue. Additionally, Facebook video watching behavior leads to meaningful insights towards optimum video length and engagement routines to fundraise massive amounts during natural disasters such as bushfires.

**Conflict of Interest**

The Authors declare no conflict of interest.